\documentclass[graphicx, 12pt]{article}
\usepackage{indentfirst}

\usepackage{graphicx}
\usepackage{caption}
\usepackage{subfigure}

\begin{document}

\small
\hoffset=-1truecm
\voffset=-2truecm
\title{\bf The tunneling radiation of Kehagias-Sfetsos black hole
under generalized uncertainty principle}

\author{Lingshen Chen\hspace {1cm} Hongbo Cheng\footnote
{E-mail address: hbcheng@ecust.edu.cn}\\
Department of Physics,\\ East China University of Science and
Technology,\\ Shanghai 200237, China\\}

\date{}
\maketitle

\begin{abstract}
We further the investigation on the Parikh-Kraus-Wilczeck
tunneling radiation of Kehagias-Sfetsos black hole under the
generalized uncertainty principle. We obtain the entropy
difference involving the influence from the inequality. The two
terms as generalizations of the Heisenberg's uncertainty promote
or retard the emission of this kind of black holes respectively.
\end{abstract}

\vspace{8cm} \hspace{1cm} PACS number(s): 04.70.Dy, 03.65.Sq, 04.62.+v\\
Keywords: Black hole evaporation, Kehagias-Sfetsos black hole, GUP

\noindent \textbf{I.\hspace{0.4cm}Introduction}

Several years ago, P. Horava proposed a UV completion of gravity
as a power-counting renormalizable theory of gravity with
anisotropic scaling of space and time like $x^{i}\longrightarrow
bx^{i}$ and $t\longrightarrow b^{z}t$ respectively and the spatial
index $i=1, 2, \cdot\cdot\cdot, D$. Here $D$ is dimentionality and
$z$ is called Lifshitz index [1, 2]. Within the frame of
Horava-Lifshitz (HL) gravity mentioned above, a kind of black hole
solution was obtained, called Kehagias-Sfetsos (KS) solution [3].
It is interesting that the KS black hole has two capture cross
sections from inner and outer horizon due to the effect from HL
scheme, but this kind of massive source becomes Schwarzschild
black hole at infinity [3]. The various contributions have been
paid to the KS black holes. According to the particle geodesics
around the KS black hole, it was found that the particles will be
either scattered or trapped near the horizon and interior space
and the coupling constant belonging to the black hole governed by
HL gravity will weaken the gravitation while modify the
trajectories of the particle considerably [4]. In the
Kraus-Parikh-Wilczek issue as Hawking radiation based on the
semi-classical tunneling [5-11], the emission rate subject to the
logarithmic entropy of KS black hole as an asymptotically flat
solution of the HL gravity was calculated [12]. This work can help
us to further understand the generalized gravity. The local
thermodynamics of KS black hole was studied and the phase
transition and stability of this kind of black hole were exhibited
[13].

During the developments of quantum gravity and black holes, some
new proposals such as the generalization of the Heisenberg's
uncertainty principle generated. The quantum gravity needs a
minimal length of the order of the Planck length [14-20]. Further
the generalized uncertainty principle (GUP) as the generalization
of Heisenberg's scheme was initiated and certainly modifies the
quantum mechanics [21]. The GUP because of the existence of the
fundamental scale of length attracted more attention [15, 22-27].
The generalization of the Heisenberg uncertainty relation in
one-dimensional space can be used to cure the divergence from
states density near the black hole horizon and relate the entropy
of black hole to a minimal length as quantum gravity scale [28,
29]. The influence from the GUP on the Beckenstein-Hawking black
hole entropy in the high-dimensional spacetime was discussed while
the black hole radiation was investigated with the help of the
tunneling formalism [28]. The Hawking tunneling radiation from
black holes under the GUP corrections was also studied in the
world that has more than four dimensions [29]. The tunneling
radiation of a black hole involving a $f(R)$ global monopole under
GUP was considered [30]. Recently a parameter belonging to the GUP
is estimated in virtue of the leading quantum corrections to the
Newtonian potential [31] and the gravitational wave event GW150914
[32] respectively. The influence from the GUP seems not to be
ignored.

It is necessary to promote the research on the tunneling radiation
of the KS black holes based on the GUP. The terms associated with
the Newtonian constant need to be added in the Heisenberg's
inequality once the gravity is taken into account [15, 22-27].
Research on the Hawking emission due to the KS black holes should
not rule out the inequality including the gravitational
corrections. As mentioned above, the tunneling radiation relating
to logarithmic entropy of KS black hole has been calculated with
Kraus-Parikh-Wilczek technique [12]. Now we plan to reexamine the
entropy and tunneling radiation of the same kind of black holes
with the same technique while the Heisenberg's uncertainty is
generalized. We wonder how the GUP modifies the entropy difference
and the tunneling probability of the KS black holes. We will list
our results in the end.

\vspace{0.8cm} \noindent \textbf{II.\hspace{0.4cm}The entropy
difference of a radiating KS black hole under generalized
uncertainty principle}

Now we start to focus on the entropy of KS black hole involving
the modified uncertainty principle. We introduce a metric of
static and spherically symmetric black hole in the deformed HL
gravity with $\lambda=1$ [3],

\begin{equation}
ds^{2}=f(r)dt^{2}-\frac{dr^{2}}{f(r)}-r^{2}(d\theta^{2}+\sin^{2}\theta
d\varphi^{2})
\end{equation}

\noindent where

\begin{equation}
f(r)=1+\gamma r^{2}-\sqrt{\gamma^{2}r^{4}+4\gamma Mr}
\end{equation}

\noindent and $\gamma=\frac{16\mu^{2}}{\kappa^{2}}$ with constant
parameters $\mu$ and $\kappa$. As a constant, $M$ is positive. In
the limit like $r\longrightarrow\infty$, the function $f(r)$
subject to the KS black hole and shown in Eq.(2) approaches the
corresponding component of Schwarzschild metric [3, 4, 13]. The
component of KS black hole like function $f(r)$ leads the outer
and inner horizons as follow [3, 12, 13],

\begin{equation}
r_{\pm}=M(1\pm\sqrt{1-\frac{1}{2\gamma M^{2}}})
\end{equation}

\noindent Here we choose the outer ones $r_{+}$ as event horizon
$r_{H}$. From Eq.(2) and (3), we obtain the Hawking temperature
[12, 13],

\begin{equation}
T_{H}=\frac{1}{8\pi}\frac{2\gamma r_{H}^{2}-1}{r_{H}(\gamma
r_{H}^{2}+1)}
\end{equation}

We proceed our discussions in the context of GUP. The Heisenberg's
uncertainty principle can be generalized within the microphysics
regime as [23, 27, 28, 33-43],

\begin{equation}
\triangle x\triangle p\geq\frac{\hbar}{2}(1-\frac{\alpha
l_{p}}{\hbar}\triangle p+\frac{(\beta
l_{p})^{2}}{\hbar^{2}}\triangle p^{2})
\end{equation}

\noindent leading,

\begin{equation}
y_{-}\leq y\leq y_{+}
\end{equation}

\noindent with the choice,

\begin{eqnarray}
y_{\pm}=(\frac{l_{p}}{\hbar}\triangle
p)_{\pm}\hspace{6.5cm}\nonumber\\
=\frac{1}{2\beta^{2}}(\alpha+\frac{2\triangle
x}{l_{p}})\pm\frac{1}{2\beta^{2}}(\alpha+\frac{2\triangle
x}{l_{p}})\sqrt{1-(\frac{2\beta}{\alpha+\frac{2\triangle
x}{l_{p}}})^{2}}
\end{eqnarray}

\noindent where $\alpha$ and $\beta$ are dimensionless positive
parameters. The Planck length is denoted as
$l_{p}=\sqrt{\frac{\hbar G}{c^{2}}}$ with velocity of light in the
vacuum $c$. The terms with the Newtonian constant $G$ provide the
inequality with the gravitational effects. Following the procedure
of Ref. [28, 29, 44], we choose,

\begin{eqnarray}
\triangle p'=\frac{\hbar}{l_{p}}y_{-}\hspace{6cm}\nonumber\\
=\frac{\hbar}{2(\beta l_{p})^{2}}(\alpha l_{p}+2\triangle
x)[1-\sqrt{1-(\frac{2\beta l_{p}}{\alpha l_{p}}+2\triangle
x)^{2}}]
\end{eqnarray}

\noindent Approximately the uncertainty in the momentum from
Eq.(8) is [44],

\begin{equation}
\triangle p'\approx\frac{\hbar}{\alpha l_{p}+2\triangle x}
\end{equation}

\noindent The combination of the GUP and the approximation (9)
helps us to estimate the distance interval containing the
gravitational corrections [44],

\begin{equation}
\triangle x'\approx\triangle x[1+\frac{(\beta
l_{p})^{2}}{2\triangle x(\alpha l_{p}+2\triangle x)}]
\end{equation}

\noindent If the $\beta$-term in the GUP disappears, no
gravitational effect will act on the distance interval no matter
whether the other term with $\alpha$ exists or not. Like Ref. [28,
29, 44], we select the corrected interval $\triangle x'$ shown in
Eq.(10) with the original size of black hole as the lower bound on
the region like $\triangle x=2r_{H}$ to obtain the Hawking
temperature of KS black hole under the GUP as follow,

\begin{eqnarray}
T'_{H}=\frac{1}{2\pi}\frac{\gamma\triangle x'^{2}-2}{\triangle
x'(\gamma\triangle x'^{2}+4)}\hspace{4.5cm}\nonumber\\
\approx T_{H}[1+\frac{2\gamma^{2}r_{H}^{4}-5\gamma
r_{H}^{2}-1}{4r_{H}(2\gamma r_{H}^{2}-1)(\gamma
r_{H}^{2}+1)(4r_{H}+\alpha l_{p})}\beta^{2}l_{p}^{2}]^{-1}
\end{eqnarray}

\noindent Although the modified Hawking temperature has something
to do with the parameters both $\alpha$ and $\beta$ from the
gravitation, the gravitational effect will disappear if the
variable $\beta$ vanishes. The Bekenstein-Hawking entropy may be
derived from the Hawking temperature by means of the following
thermodynamics relation [5-7, 11],

\begin{equation}
T_{H}=\frac{dE}{dS}\approx\frac{dM}{dS}
\end{equation}

\noindent where $E$ is the total energy. In the case of GUP, the
Hawking temperature $T_{H}$ can be replaced as $T_{H}'$. According
to Eq.(12) and GUP, we perform the burden derivation to obtain the
corrected entropy difference of the radiating KS black hole,

\begin{eqnarray}
\triangle S=\triangle S(\gamma, \alpha, \beta)\hspace{8cm}\nonumber\\
=\pi(r_{H}'^{2}-r_{H}^{2})+\frac{3\pi C}{8}(\beta
l_{p})^{2}\ln\frac{2\gamma r_{H}'^{2}-1}{2\gamma
r_{H}^{2}-1}\hspace{4.5cm}\nonumber\\
+\frac{3\pi\gamma C}{16\sqrt{2\gamma}}(\alpha l_{p})(\beta
l_{p})^{2}(\ln\frac{\sqrt{2\gamma}r_{H}-1}{\sqrt{2\gamma}r_{H}+1}
-\ln\frac{\sqrt{2\gamma}r_{H}'-1}{\sqrt{2\gamma}r_{H}'+1})\hspace{2cm}\nonumber\\
+\pi C[\frac{(1-6C)\gamma}{64}(\alpha
l_{p})^{2}+\frac{1+6C}{8}-\frac{2}{\gamma(\alpha
l_{p})^{2}}](\beta l_{p})^{2}\ln\frac{4r_{H}'+\alpha
l_{p}}{4r_{H}+\alpha l_{p}}\nonumber\\
+\frac{2\pi}{\gamma}[1+(\frac{1}{(\alpha
l_{p})^{2}}-\frac{\gamma}{8})C(\beta
l_{p})^{2}]\ln\frac{r_{H}'}{r_{H}}\hspace{4cm}\nonumber\\
+\frac{\pi}{2\gamma}C(\frac{\gamma\alpha l_{p}}{8}-\frac{1}{\alpha
l_{p}})(\frac{1}{r_{H}}-\frac{1}{r_{H}'})(\beta
l_{p})^{2}\hspace{4cm}
\end{eqnarray}

\noindent where,

\begin{equation}
C=\frac{8}{\gamma(\alpha l_{p})^{2}-8}
\end{equation}

\noindent and $r_{H}'=M'(1+\sqrt{1-\frac{1}{2\gamma M'^{2}}})$
which is similar to Eq.(3) while $M'=M-\hbar\omega$. Here $\omega$
is a shell of energy moving along the geodesics in the spacetime
described by metric (1) [11]. In the similar process that we only
set $\beta=0$, the entropy change above return to be the results
of Ref. [12],

\begin{equation}
\triangle
S_{0}=-\pi(r_{H}^{2}-r_{H}'^{2})-\frac{2\pi}{\gamma}\ln\frac{r_{H}}{r_{H}'}
<0
\end{equation}

\noindent In the case that $\beta>0$, the parameter $\alpha$ can
amend the change in addition to the $\beta$ terms in the
Expression (13). The Hawking radiation as semiclassical quantum
tunneling for KS black hole has been investigated with the help of
Kraus-Parikh-Wilczek methodology while governed by the
Heisenberg's uncertainty principle of quantum mechanics [12]. The
HL gravity that has a kind of solution named as KS black hole
could be a candidate of quantum gravity which needs a minimal
length of the order of the Planck length and initiates the GUP
[14-21]. We reinvestigate the emission rate $\Gamma=e^{\triangle
S}$ [10, 11] and entropy of KS black hole under the GUP. In order
to explore the influence from GUP on the quantum tunneling rate of
the radiating KS black hole, we discuss the ratio $\frac{\triangle
S}{\triangle S_{0}}$ versus the variables $\alpha$ and $\beta$.
The two parts with $\alpha$ and $\beta$ respectively making up the
corrections contribute the opposite effects to the emission
spectrum. The figure 1 declares that the bigger magnitude of
$\alpha$ will decrease the absolute value of the entropy change,
which stimulates the emission rate of the KS black holes. It is
found that the stronger influence from $\beta$ part will increase
the absolute value of $\triangle S$ much more greatly in figure 2.
The nature of entropy difference of this kind of black hole is
negative, so greater absolute value of $\triangle S$ retards the
emission rate. The figures both demonstrate that the stronger HL
coupling $\gamma$ makes the whole curve of ratio $\frac{\triangle
S}{\triangle S_{0}}$ up, or equivalently make the radiation
slower.

\vspace{0.8cm} \noindent \textbf{III.\hspace{0.4cm}Discussion}

In this work we study the entropy difference of radiating KS black
hole under the generalized uncertainty principle. The GUP has two
kinds of corrections. The negative part shown as $\alpha$-term
makes the entropy change large, which promotes the radiation of KS
black hole to keep it stable. The positive one like $\beta$-term
leads absolute value of entropy difference $\triangle S$ whose
nature is negative to be much larger if this term is greater,
which means that the positive part added in the GUP will damp the
emission of KS black holes. The stronger HL coupling $\gamma$ from
the deformed gravity makes the ration $\frac{\triangle
S}{\triangle S_{0}}$ greater, leading the black hole radiation
slower.

\vspace{1cm}
\noindent \textbf{Acknowledge}

This work is supported by NSFC No. 10875043.

\newpage
\begin{figure}
\setlength{\belowcaptionskip}{10pt} \centering
\includegraphics[width=15cm]{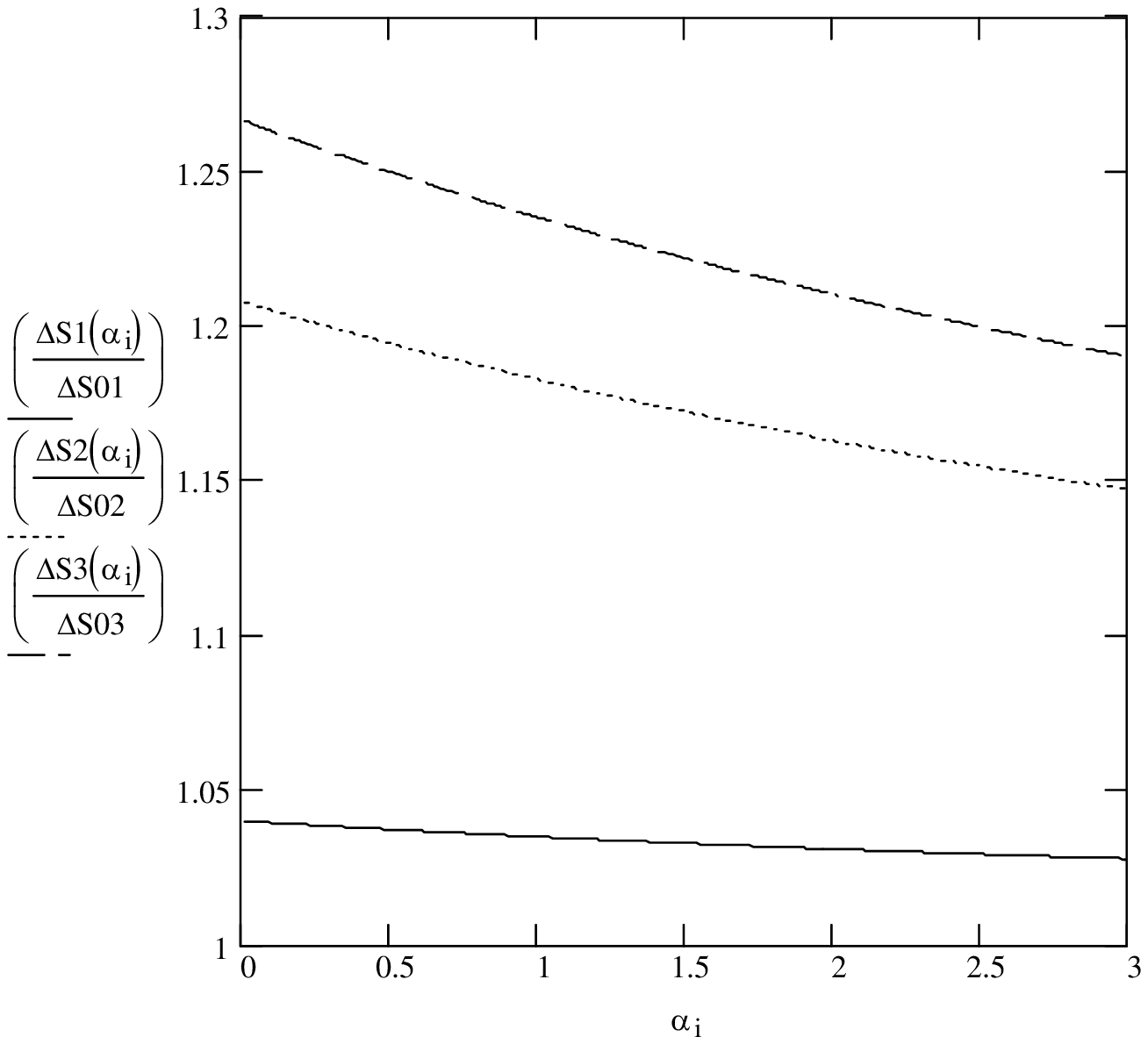}
\caption{The solid, dashed, dot curves of the dependence of the
ratio $\frac{\bigtriangleup S}{\bigtriangleup S_{0}}$ on $\alpha$
for $\gamma=1, 1.1, 1.2$ respectively with $\beta=5$ and $l_{p}=1$
for simplicity.}
\end{figure}

\newpage
\begin{figure}
\setlength{\belowcaptionskip}{10pt} \centering
\includegraphics[width=15cm]{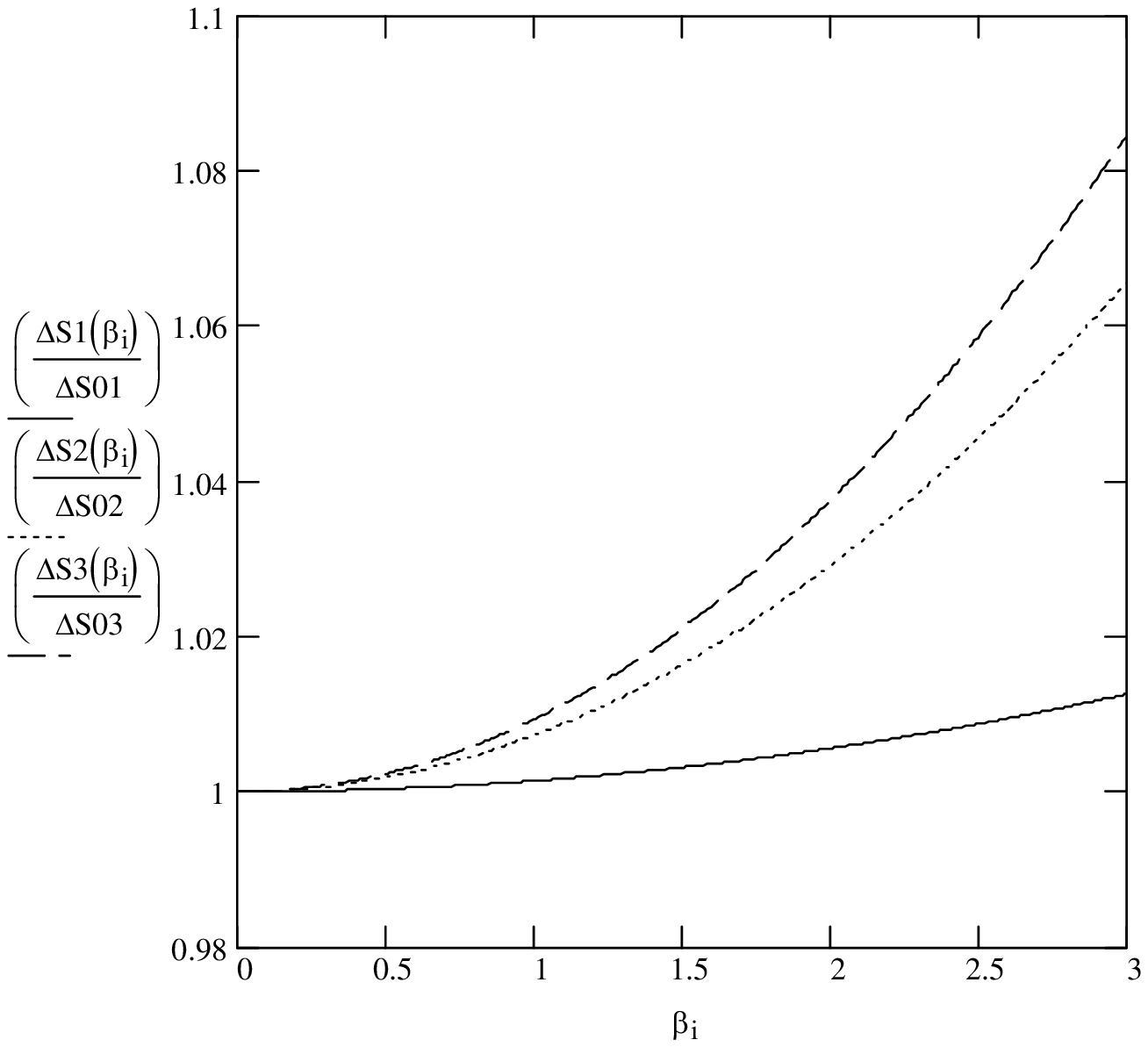}
\caption{The solid, dashed, dot curves of the dependence of the
ratio $\frac{\bigtriangleup S}{\bigtriangleup S_{0}}$ on $\alpha$
for $\gamma=1, 1.1, 1.2$ respectively with $\beta=5$ and $l_{p}=1$
for simplicity.}
\end{figure}

\end{document}